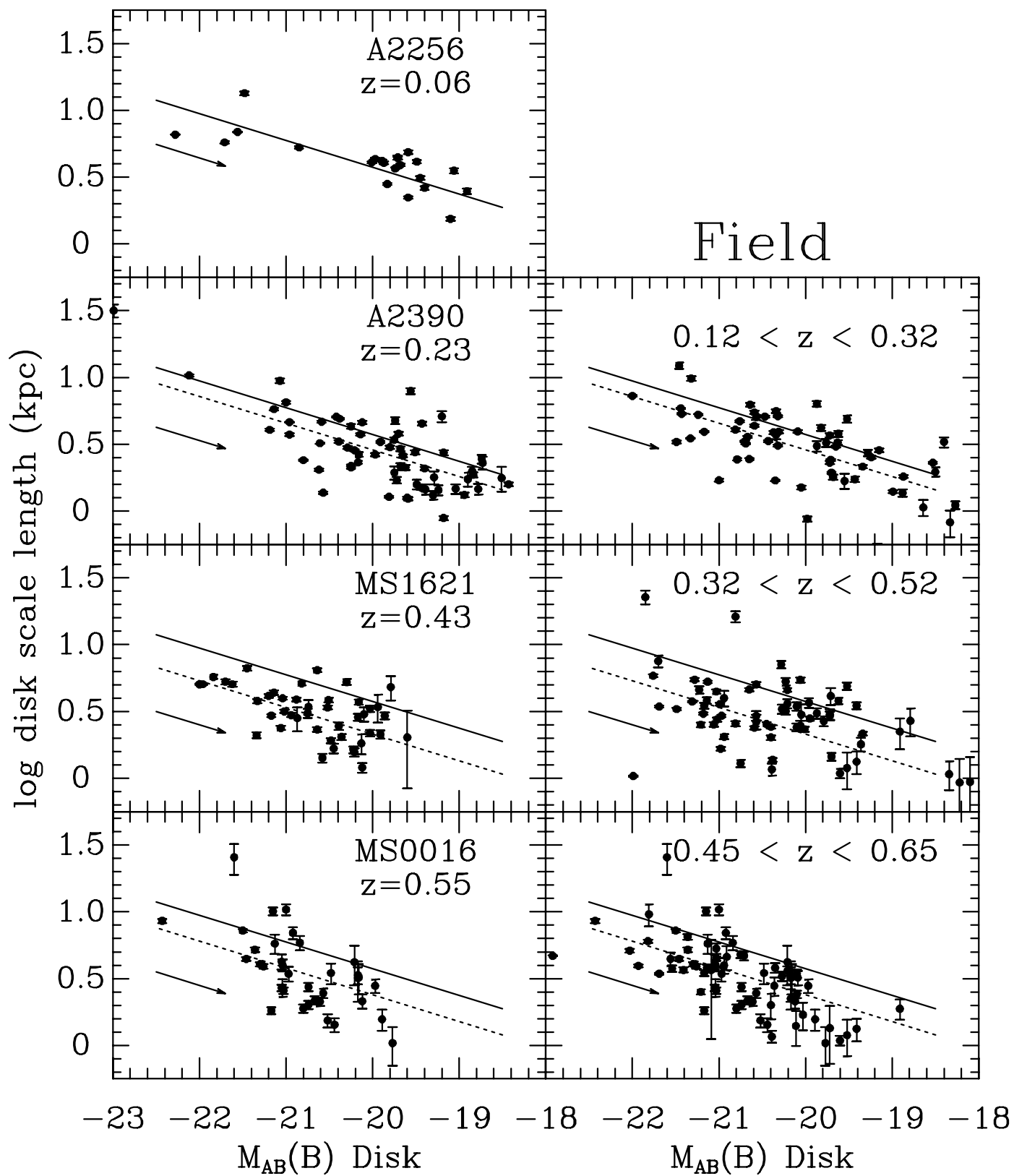

# Evolution of galactic disks in clusters and the field at $0.1 < z < 0.6$ in the CNOC survey

David Schade[1], R. G. Carlberg[1], H.K.C. Yee[1] and Omar López-Cruz[2]

*Department of Astronomy, University of Toronto, Toronto, Canada M5S 3H8*

Erica Ellingson[1]

*Center for Astrophysics and Space Astronomy, University of Colorado, CO 80309, USA*

## ABSTRACT

Two-dimensional surface photometry is presented for a sample of 351 late-type galaxies with $0.12 < z < 0.65$. These objects are drawn from the Canadian Network for Observational Cosmology (CNOC) cluster survey and are either spectroscopically confirmed members of clusters at $z = 0.23$ (64 galaxies), 0.43 (45), and 0.55 (36) or field galaxies with similar redshifts. Galaxies in the rich cluster Abell 2256 at $z = 0.06$ were also analyzed with the same methods to provide a local reference point. At redshifts of (0.23, 0.43, 0.55) the disk surface brightness in cluster late-type galaxies is higher in the $B$-band by $\Delta\mu_0(B) = (-0.58 \pm 0.12, -1.22 \pm 0.17, -0.97 \pm 0.2)$ mag, respectively, relative to the Freeman (1970) constant surface-brightness relation; whereas disks in cluster galaxies at $z = 0.06$ are consistent with that relation. Field galaxies show a progressive disk-brightening with redshift that is consistent with that seen in the cluster population. Taken together with similar measurements of early-type galaxies (Schade et al. 1996a), these results suggest that the evolution of the field and cluster galaxy populations are similar, although we emphasize that our sample of cluster galaxies is dominated by objects at large distances (up to 3 Mpc) from the dense cluster core, so that the implications of these findings with respect to the Butcher-Oemler effect and the morphology-density relation will not be clear until an analysis of galaxy properties as a function of cluster-centric distance is completed.

*Subject headings:* galaxies:evolution—galaxies:fundamental parameters

---

## 1. INTRODUCTION

Galactic disks exhibit an apparent diversity of star-formation histories. Kennicutt, Tamblyn, & Congdon (1994) show that early-type disks formed stars more vigorously long ago whereas late-type objects have present-day star-formation rates comparable to their past average values. Models using exponentially-declining star-formation rates to reproduce present-day disk colors (Charlot & Bruzual 1991) require e-folding time scales of 3 and 9 Gyr for early- and late-type disks respectively. Disk evolution is thus believed to continue to the present day with ongoing infall of gas (e.g., Gunn 1987). Because they form the bulk of the luminous population (e.g., Buta et al. 1994), disk galaxies must play an important role in the overall evolution of the galaxy population (King & Ellis 1985).

Deep number counts of galaxies (e.g., Tyson 1988) have been interpreted as evidence of galaxy evolution (Lilly 1993, Ellis 1987) although inferences about the physical nature of this evolution have necessarily relied upon detailed modeling. Many of the ambiguities in interpretation of the counts have been erased with the completion of deep redshift surveys (Lilly et al. 1995, Glazebrook et al 1995) which allow a direct construction of the evolving galaxy luminosity function (Lilly et al. 1995; Ellis et al 1996). The luminosity function (LF) of field galaxies from the Canada-France Redshift Survey (CFRS) (Lilly et al. 1995) reveals significant evolution in the population of galaxies bluer than present-day $Sbc$. Between $z \sim 0.3$ and $z \sim 0.75$ this evolution can be characterized as an increase in space density by a factor of 3, or as a uniform increase in the luminosity of these galaxies by $\sim 1$ mag, or, of course, as part of some more complex, evolutionary scenario. At lower redshifts, Ellis et al (1996) detect evolution in the galaxy LF between $z = 0.1$ and $z = 0.5$ which they interpret as predominantly a steepening of the faint-end slope of the LF. The luminosity function represents the statistical properties of the galaxy population and cannot directly reveal the evolutionary behavior of individual galaxies.

Schade et al. (1995) used HST imaging of a sample of 32 galaxies from the CFRS to determine physical properties of individual objects at $0.5 < z < 1.2$. Of the blue galaxy population, half are disk-dominated galaxies that are apparently normal except that their mean $B$-band surface brightness is higher by $1.2 \pm 0.25$ mag than the local Freeman (1970) value. One-third of the blue galaxies are dominated by compact blue components (the "blue-nucleated galaxies"), and show a high frequency of asymmetric/peculiar structure. These results are confirmed by Schade et al. (1996b) for a sample of 143 galaxies with ground-based imaging. Evidently most late-type galaxies in these samples had significantly higher rates of star-formation at $z \sim 0.75$ than they do at the present time.

The role played by disk-dominated galaxies in the evolution of the cluster population is not clear. The fraction of blue galaxies near the centers of clusters increases with redshift (Butcher & Oemler 1984) and these blue galaxies appear similar to normal late-type systems, perhaps with a high incidence of disturbed morphology (Dressler et al 1994). The morphology-density relation (Dressler, Gunn, & Schneider 1985, Whitmore, Gilmore, & Jones 1993) shows that although ellipticals dominate the population in cluster cores (comprising more than 50% of the population at $R < 0.25$ Mpc), the mix of morphological types rapidly approaches the mix in the field for $R > 1$ Mpc.

The connection between evolving cluster and field galaxies is a critical issue for two reasons. First, the influence of environment on the evolution of cluster galaxies cannot be understood without an understanding of the evolutionary state of field galaxies at the same redshifts. Second, the problem of determining $\Omega_\circ$ through the cluster mass-to-light ratio (Carlberg et al. 1996) is strongly coupled to the problem of differential evolution between clusters and the field because galaxy luminosities are used to normalize the cluster mass to an equivalent volume in the field. Differential evolution between the cluster and field populations would, without correction, result in a systematic error in the $\Omega_\circ$ measurement. The Canadian Network for Observational Cosmology (Carlberg et al. 1994, Yee, Ellingson, & Carlberg 1996) cluster survey contains redshifts for 2600 galaxies at $0.17 < z < 0.55$ evenly distributed between clusters and the field and thus provides a unique opportunity to address this issue (see also Schade et al. 1996a).

This *Letter* concentrates on the morphological analysis of 351 *late-type* cluster and field galaxies at $0.1 < z < 0.6$ representing the first phase of a comprehensive study of the evolution of cluster galaxies and differential evolution of the cluster and field populations. Observations, and procedure are described in §2. The relation between size and luminosity is presented in §3 and the results are discussed in §4. It is assumed throughout this paper that $H_o = 50$ km sec$^{-1}$ Mpc$^{-1}$ and $q_o = 0.5$.

## 2. OBSERVATIONS AND PROCEDURE

Imaging for three CNOC cluster fields was obtained in June and October 1993 using the Canada-France-Hawaii Telescope Multi-Object Spectrograph (MOS). Details of fields, seeing, integration times, and the measuring procedure are given in Schade et al. (1996a). The two-dimensional luminosity distributions of all galaxies with cataloged velocities were measured by fitting two-component ($r^{1/4}$ bulge and exponential disk) models (described by Schade et al. 1996b) to the $r$-band MOS imaging. These models are integrated over each pixel, convolved with the empirical point-spread function, and compared to the data. The fitted data were images that had been "symmetrized" (see Schade et al. 1995) to minimize the effects of neighbors and asymmetric structure. A modified Levenberg-Marquardt algorithm was used to locate the minimum in $\chi^2$ which corresponds to the best-fit model. The best-fit models yield size, surface brightness, and axial ratio information for each component, as well as the fractional bulge luminosity ($B/T$). In addition to the 561 galaxies in 3 CNOC fields (with a 4% fit failure rate), fits were done to 100 bright galaxies on an 1800 second B image of the cluster A2256 from the López-Cruz & Yee survey (López-Cruz 1996) obtained with the Kitt Peak 0.9-meter telescope. These galaxies were chosen to be within $\pm 0.1$ magnitudes of the tight red-galaxy sequence in the $B$ vs. $B - R$ diagram. Because of the stringent color selection there is no guarantee that these are typical disk galaxies.

Galaxies with a bulge fraction $B/T < 0.5$ as measured from the best-fit models were defined as late-type or disk-dominated galaxies. The distinction between field and cluster galaxies was made strictly on the basis of velocity adopting the cluster redshift limits given in Table 1 of Carlberg (1996). It is important to note that the cluster population as defined here is dominated by galaxies at large distances (up to 3 Mpc) from the cluster core.

The observed $g - r$ colors were used to transform the $r$-band magnitudes into restframe $M_{AB}(B)$ luminosities ($B_{AB} = B - 0.17$) and colors $(U - V)_{AB}$ [$(U - V)_{AB} = (U - V) + 0.7$] based on interpolation among the spectral energy distributions of Coleman, Wu, and Weedman (1980) as described by Lilly et al. 1995. The galaxies in A2256 were also K-corrected according to Coleman, Wu, and Weedman (1980).

## 3. RESULTS

Figure 1 shows the relation between $\log h$ (disk scale length in kiloparsecs) and disk luminosity (in the $AB$ system) for cluster and field galaxies at $0.06 < z < 0.6$. Bulge contributions to the luminosity have been removed using the best-fit galaxy model parameters although this correction is negligible for the majority of galaxies and has no effect on the global trends measured here. A constant surface-brightness slope of $\Delta M/\Delta \log h = -5.0$ (close to the the mean value $-5.3 \pm 0.5$ of fits to all of the cluster and field loci individually) was adopted for the cluster and field galaxies. Measuring the shifts in the $M_B - \log h$ plane is thus equivalent to measuring the change in mean central surface brightness of the sample. We adopt the Freeman (1970) value $\mu_{0,AB}(B) = 21.48$ as the local ($z = 0$) fiducial value.

Plotted on Figure 1 is the Freeman relation (solid line) along with the best-estimate lines for the galaxy loci for each cluster (dotted lines). These *cluster* loci are superimposed on the field galaxy panels (dotted lines; these are *not* the best-fit *field* galaxy lines). The change in mean surface brightness for the clusters and field along with their uncertainties are given in Table 1. The data in Figure 1 and Table 1 show that the disks of late-type galaxies have progressively higher mean surface brightness (relative to the Freeman (1970) law) with increasing redshift. The amount of brightening seen in the clusters is indistinguishable from that in the field. Because surface brightness depends only on redshift the derived evolution is independent of cosmology.

## 4. DISCUSSION

These observations lead to two main conclusions. First, the disks of late-type galaxies in clusters show a progressive increase in surface brightness with redshift such that disks at $z = 0.55$ have a mean surface brightness $\sim 1$ mag brighter than the Freeman (1970) value for local galaxies. Second, the evolution in surface brightness of field disk galaxies is indistinguishable from that observed in the clusters.

A recent discussion of the validity of the Freeman law is given by de Jong (1996) who concludes that the law needs modification because there is no single preferred value of the central surface brightness in galactic disks but that there does exist an upper limit. That dataset also leads to the conclusion that galaxies with low surface brightness (relative to the Freeman law) are generally of late Hubble type and of lower luminosity (see Figures 3 and 6 of de Jong 1996). If one were to select a sample from the de Jong dataset that had a luminosity distribution similar to the CNOC sample discussed here, one would, in fact, recover a distribution of central surface brightness very similar to the Freeman law (see Figure 8 and Table 1 of de Jong 1996). This fact argues that the comparison with the Freeman law is reasonable.

The upper limit to disk central surface brightness (at $\mu_o(B) \sim 20$ mag arcsec$^{-2}$) claimed by de Jong (1996) provides a way to check our derived evolution. The arrows plotted on Figure 1 are segments of the lines defining this upper limit on central surface brightness *assuming that the upper limit evolves in the same manner as the mean central surface brightness* (given in Table 1). It can be seen that evolution in the upper limit is *required* relative to the no-evolution line shown in the A2256 panel, and that the evolution in the upper limit is consistent with the evolution in the mean surface brightness. This fact argues that our derived evolution is not produced by selection effects working to exclude low-surface brightness galaxies.

Figure 2 shows the derived values of disk-brightening as a function of redshift for the present sample of cluster and field galaxies. Also shown are measurements from the Canada-France Redshift survey (Schade et al. 1995,1996b) which extend the field-galaxy relation to $z = 0.75$. Lilly et al. (1996) show that the global star-formation rate at $z \sim 1$ (including galaxies of all morphological types and colors) was higher by a factor of 7-15 (depending on cosmology) than the present-day rate and that the $B$-band luminosity density increases as $L \propto (1+z)^{2.7\pm0.5}$ (for the cosmology adopted here). We plot this function on figure 2. It is important to note that there is no normalization involved in this plot. It simply demonstrates that the *shape* of the variation of global luminosity density with redshift is similar to the shape of the variation of disk surface brightness with redshift.

The results reported here imply that the majority of luminous disks have enhanced surface brightness and, thus, increased luminosity at higher redshifts. The luminosity functions from the CFRS (Lilly et al. 1995) show clear evidence of evolution among the blue population of galaxies (mostly disks) between $0.2 < z < 0.5$ and $0.5 < z < 1$. This is consistent with luminosity evolution of individual objects and, in fact, morphological measurements of disks at $0.5 < z < 1.2$ show an increase in mean surface brightness similar to that reported here. There is little overlap between the CFRS and CNOC samples in the $M - z$ plane so that a direct comparison at lower redshift is not possible. Ellis et al. (1996) detect evolution in the LF between $z \sim 0.1$ and $z \sim 0.5$ that is consistent with the present results (comparing the appropriate ranges in luminosity). On the other hand, they see no sign of evolution (at the luminosities of interest here) between $z = 0$ and $z = 0.25$ where we would expect $\sim 0.5$ mag of brightening. The meaning of these comparisons between the Ellis et al. (1996) sample and the CNOC results is uncertain because of the large difference in selection bands ($B$ versus Gunn-$r$ for CNOC). Thus, there is general agreement (Lilly et al. 1995, Ellis et al. 1996) that the luminosity function shows evolution beginning at low redshift ($z \sim 0.3$). The results presented here are broadly consistent with the available luminosity function results.

A more detailed study (Schade et al. in preparation) will include an analysis of the scatter in the $M_{AB}(B) - \log h$ relations and an examination of the correlations between surface brightness enhancement, luminosity, color, and spectral indices. The present results have broad implications for field galaxy evolution as well as cluster galaxy populations, e.g., the Butcher-Oemler (1984) effect and the morphology-density relation (Dressler 1985), but we emphasize, once again, that our cluster galaxy samples are dominated by objects far from the cluster core so that direct comparisons with cluster studies that concentrate on the core regions are not possible at this stage. A comprehensive study of the relation between galaxy properties and distance from the core is the primary goal of the CNOC morphology project.

In this paper and earlier ones (Schade et al. 1995, Schade et al. 1996a, Schade et al. 1996b, Barrientos, Schade, & López-Cruz 1996) significant evolution has been reported in the relations between size, luminosity and surface brightness for the population of luminous galaxies over the redshift range $0 < z < 1$. No significant difference has been found between the evolutionary properties of field and cluster galaxies. If interpreted as luminosity evolution of individual galaxies, the evolution of ellipticals is consistent with models where the bulk of the stellar population is formed in a single burst early in the history of the universe and evolves passively thereafter (e.g., Tinsley 1972; Bruzual 1993) whereas disk systems appear to require star-formation rates that were significantly higher at $0.5 < z < 1.0$ than at the present time.

We thank Simon Lilly, Felipe Barrientos, and Pedro Colin for helpful discussions. This work was supported financially by NSERC of Canada.

# REFERENCES

Barrientos, F., Schade, D., & López-Cruz, O. 1996 ApJ, April 1 in press

Buta, R., Mita, S., de Vaucouleurs, G., & Corwin, H. 1994, AJ, 107, 118

Butcher, H. & Oemler, A. 1984, ApJ, 285, 423

Bruzual, G. 1993, ApJ, 405, 538

Carlberg, R. G., Yee, H. K. C., Ellingson, E., Abraham, R., Gravel, P., Morris, S., & Pritchet, C. J. 1996, ApJ, May 1

Carlberg, R. G., Yee, H. K. C., Ellingson, E., Pritchet C., Abraham, R., Smecker-Hane, T., Bond, J. R., Couchman, H. M. P., Crabtree, D., Crampton, D., Davidge, T., Durand, D., Eales, S., Hartwick, F. D. A., Hesser, J. E., Hutchings, J. B., Kaiser, N., Mendes de Oliveira, C., Myers, S. T., Oke, J. B., Rigler, M. A., Schade, D., & West, M. 1994, JRASC, 88, 39

Charlot, S., & Bruzual, A. G. 1991, ApJ, 367 126

Dressler, A., Oemler, A., Sparks, W., & Lucas, R. 1994, ApJ, 435, L23

Dressler, A., Gunn, J., & Schneider, D. 1985, ApJ, 294, 70

Ellis, R., Colless, M., Broadhurst, T., Heyl, J., & Glazebrook, K. 1996, preprint

Ellis, R. 1987, in *Observational cosmology; Proceedings of the IAU Symposium Beijing* Dordrecht, D. Reidel Publishing Co., p. 367

Freeman, K. 1970, ApJ, 160, 811

Glazebrook, K., Ellis, R., Colless, M., Broadhurst, T., Allington-Smith, J., & Tanvir, N. 1995, MNRAS, 273, 157

Gunn, J.E. 1987 in *The Galaxy: Proceedings of the NATO Advanced Study Institute* ed. Gilmore, G. & Carswell, B. Dordrecht, D. Reidel Publishing Co., p. 413-429.

de Jong, R. 1996 preprint, to appear in Astronomy and Astrophysics

Kennicutt, R., Tamblyn, P., & Congdon, C. 1994, ApJ, 435, 22

King, C., & Ellis, R. 1985, ApJ, 288, 456

Lilly, S.J., Le Fevre, O., Hammer, F., Crampton, D., 1996, ApJ, in press

Lilly, S.J., Tresse, L., Hammer, F., Crampton, D. & Le Fèvre, O. 1995, ApJ, 455, 108

Lilly, S. J. 1993, ApJ, 411, 501

López-Cruz 1996, PhD Thesis, University of Toronto

Schade, D., Lilly, S., Crampton, D., Le Fèvre, O., Hammer, F., & Tresse, L. 1995, ApJ, 451, L1

Schade, D., Carlberg, R.G., Yee, H.K.C., López-Cruz, O., & Ellingson, E. 1996a, ApJ, submitted

Schade, D., Lilly, S., Le Fèvre, O., Hammer, F., & Crampton, D. 1996b, ApJ, in press

Tinsley, B. 1972, ApJ, 178,319

Tyson, J. A. 1988, AJ, 96, 1

Whitmore, B., Gilmore, D., & Jones, C. 1993, ApJ, 407, 489

Yee, H.K.C., Ellingson,E., & Carlberg, R. G., ApJS, 102, 269

Fig. 1.— The relation between disk luminosity and scale length for late-type ($B/T < 0.5$) galaxies at $0.06 < z < 0.55$. The left panel shows cluster galaxies and the right panel shows galaxies in the field at each corresponding redshift. Disk surface brightness (or equivalently the change in luminosity) shifts progressively toward brighter values with redshift. The best-fit constant surface brightness relations for cluster galaxies are plotted as dotted lines and these *cluster* fits are superimposed on the field data. The local Freeman (1970) constant surface brightness relation is shown as a solid line in each panel. Arrows indicate the expected evolution of the upper limit in central surface brightness (see text).

Fig. 2.— The enhancement of disk surface-brightness (excess above the Freeman law) is shown as a function of redshift. Cluster galaxies (circles) and field galaxies (squares) are from the CNOC survey. Also shown are field galaxies (triangles) from the Canada-France Redshift Survey (Schade et al. 1995,1996b) at $z \sim 0.75$. Also plotted is a function describing the evolution in the global luminosity density $L_B \propto (1+z)^{2.7}$ derived by Lilly et al. (1996).

Table 1
Evolution of the disk surface brightness of CNOC galaxies

| Cluster | z | $\Delta\mu_0(B)$ | N | Field | $\Delta\mu_0(B)$ | N |
|---|---|---|---|---|---|---|
| Abell 2390 | 0.228 | $-0.58 \pm 0.12$ | 64 | $0.12 < z < 0.32$ | $-0.55 \pm 0.13$ | 62 |
| MS1621+26 | 0.427 | $-1.22 \pm 0.17$ | 45 | $0.32 < z < 0.52$ | $-0.95 \pm 0.15$ | 71 |
| MS0016+16 | 0.547 | $-0.97 \pm 0.20$ | 36 | $0.45 < z < 0.65$ | $-1.06 \pm 0.12$ | 73 |

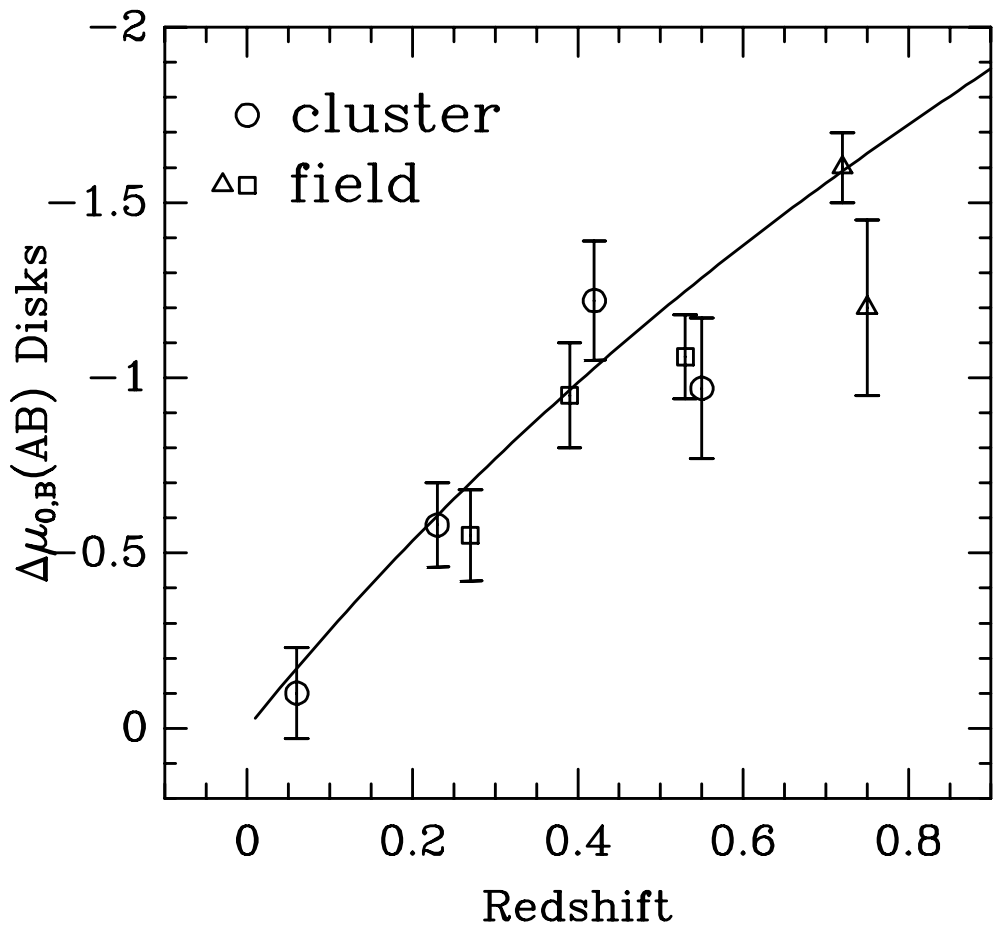